# Quantitative Multiscale Analysis using Different Wavelets in 1D Voice Signal and 2D Image

Niraj Shakhakarmi

Department of Electrical & Computer Engineering, Prairie View A&M University (Texas A&M University System)
Prairie View, Houston, Texas, 77446, USA

### Abstract

Mutiscale analysis represents multiresolution scrutiny of a signal to improve its signal quality. Multiresolution analysis of 1D voice signal and 2D image is conducted using DCT, FFT and different wavelets such as Haar, Deubachies, Morlet, Cauchy, Shannon, Biorthogonal, Symmlet and Coiflet deploying the cascaded filter banks based decomposition and reconstruction. The outstanding quantitative analysis of the specified wavelets is done to investigate the signal quality, mean square error, entropy and peak-to-peak SNR at multiscale stage-4 for both 1D voice signal and 2D image. In addition, the 2D image compression performance is significantly found 93.00% in DB-4, 93.68% in bior-4.4, 93.18% in Sym-4 and 92.20% in Coif-2 during the multiscale analysis.

**Keywords:** *Quantitative, Multiscale Analysis, Different Wavelets, One Dimensional Voice Signal, Two Dimensional Image.*

## 1. Introduction

First generation wavelets transform essentially needs the Fourier transform and the basis functions which are dyadically scalable with translation property of one particular mother basis function. These are the first non-trivial wavelets developed around 1980s. These include the Daubechies wavelet, Haar wavelet, Shannon Wavelet, Coiflets Wavelet and the Meyer wavelet. The major drawback of the first generation wavelet is that it can be deployed for infinite or periodic signals and cannot be optimized in the bounded domain. These wavelets transforms (WTs) are used in identifying pure frequencies, de-noising signals, detecting discontinuities and breakdown points, detecting self-similarity and compressing images.

Second generation wavelets transform originates with concept of Lifting scheme to maintain the time-frequency localization and fast algorithms instead of fourier domain to deploy in geometrical applications. This should replace translation and dilation as well as any Fourier analysis. The basic algorithm of the lifting scheme, is to split up even samples then are adjusted to serve the coarse version of the original signal data in even set and odd set dilation as well as any Fourier analysis. The basic algorithm of the lifting scheme is to split even samples then are adjusted to serve the coarse version of the original signal data in even set and odd set then predict odd signal using even part to detect the missing parts called details and update even samples for adjustment to serve the coarse version of the original signal. These WTs are extensively used for lossy data compression, in geographical data analysis, computer graphics and efficient coding in compression algorithm.

Third generation wavelets transform are the complex wavelet transform (CWT) with the complex-valued extension to the standard discrete wavelet transform (DWT). It is typically two-dimensional wavelet transform deployed for the multi-resolution, sparse representation, and useful feature characterization based on the structure of an image. The major pros are that these WTs do not exhibit oscillations, lack of directivity, aliasing and degree of shift-variance in its magnitude. But, the major cons are that it exhibits two dimension of the signal being transformed and yields the redundancy compared to a separable.

Next generation wavelets transform optimize the PSNR, error free, lossless and advanced multi level resolution. These wavelets will be more advanced in terms of efficiency and performance. These WTs are still under research and they will focus specific applications such as human vision characterization, frequency localization, feature extraction, seismic analysis, bio-medical analysis and so on.

Multiscale analysis represents the hierarchy of structural implementation to enhance the physical characteristics of the signal (both 1D and 2D). When the multiscale stage (level) is increased then it provides the fine resolution from coarse resolution. In other words, it is the systematic process to analyze signal at lower multiscale stage with coarse resolution and then higher multiscale stage with fine resolution [1-2]. Thus, higher stage of the multiscale using wavelets provides significant multiresolution improving signal quality. This is deployed using different kinds of wavelets in signal decomposition and reconstruction to investigate their performance at stage-3 and stage-4. The Haar wavelet, Deubechies wavelet, Morlet wavelet, Cauchy wavelet, Shannon wavelet, DCT, FFT, Biorthogonal wavelet, Symmlet wavelet and Coiflet wavelet are deployed in 1D signal and 2D image in this paper.

## 2. Problem & Proposed Solution

The problem is to analyze and compare the quantitative mutiscale features of different wavelets transform and determine the qualitatively suitable wavelets on 1D voice signal and 2D image for multi-resolution. This is addressed by decomposition and reconstruction of 1D voice signal and 2D image by deploying different wavelets transform at the third and fourth multi-resolution stage. In addition, the quantitative analysis of 1D signal and 2D image is done in terms of SNR, MSE, entropy and PSNR for different wavelets transform.

The proposed solution includes the following aspects:

- 1D Signal decomposition and reconstruction at stage-4 using different wavelets

- Quantitative analysis of 1D Signal at stage-4

- 2D Image decomposition and reconstruction at stage-4 using different wavelets

- Quantitative analysis of 2D Image at stage-4

- 2D Image compression at stage-4

### 2.1 One Dimensional Signal Decomposition & Reconstruction

One dimensional discrete wavelet transform is used for 1 D signal decomposition and reconstruction in time-scale (frequency) representation of non-stationary signals. It is based on multi-resolution approximation in which a function uses scaling function at various resolutions so that the lost details can be recovered using wavelet functions and the original signal is reconstructed by adding approximation and detail coefficient. It is deployed by a sequence of low pass and high pass filters [3-6]. Low pass (LP) filters are associated with the scaling function and provide approximation whereas high pass (HP) filters are associated with the wavelet function and provide detail lost in approximating the signal.

### 2.1.1 1D Signal Analysis (Decomposition) at Stage-4 using Different Wavelets

1D signal decomposition is done using a sequence of LP and HP filter banks at four different stages by cascading at LP downsample decimated by 2 as shown in Fig-1.

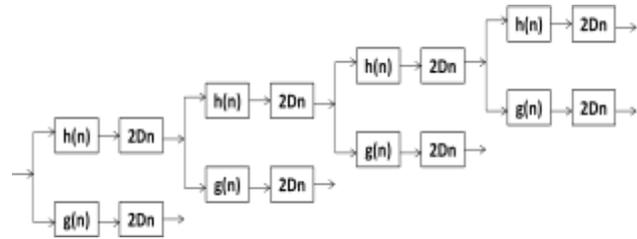

Fig. 1 1D Signal Decomposition at stage- 4

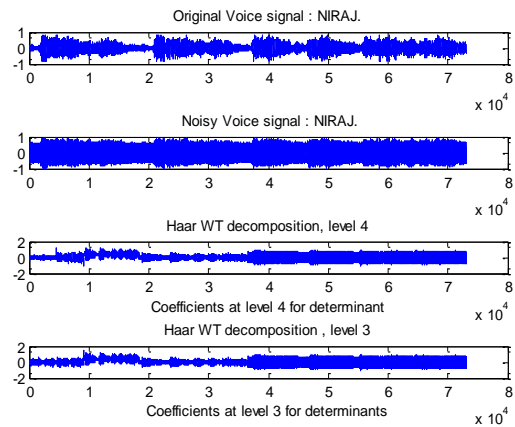

Fig. 2 Haar WT Decomposition at stage- 4

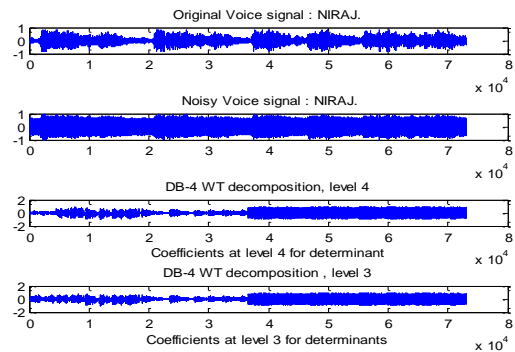

Fig. 3 DB-4 WT Decomposition at stage- 4

1D voice signal is decomposed at stage-3 & stage-4 using DCT, FFT and different wavelets such as Haar, Deubachies, Morlet, Cauchy, Shannon, Biorthogonal, Symmlet and Coiflet as illustrated in Fig. 2 to Fig. 12.

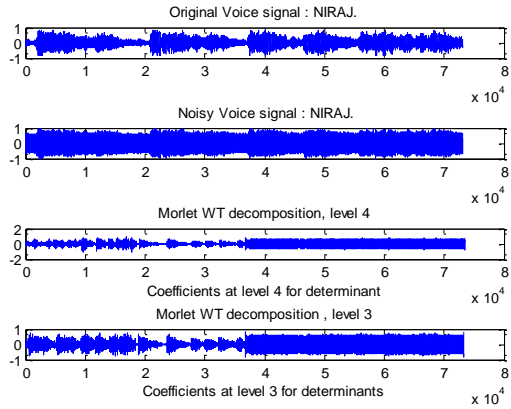

Fig. 4   Morlet WT Decomposition at stage- 4

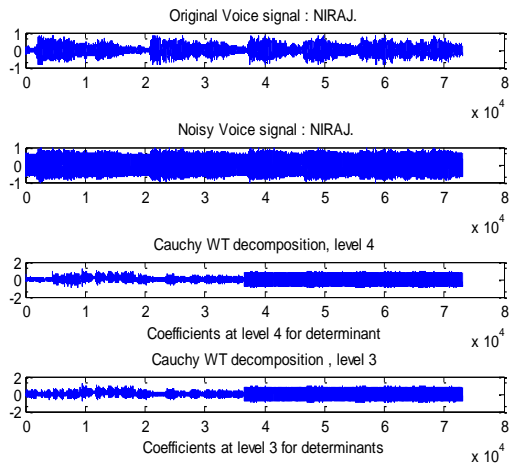

Fig. 5   Cauchy WT Decomposition at stage- 4

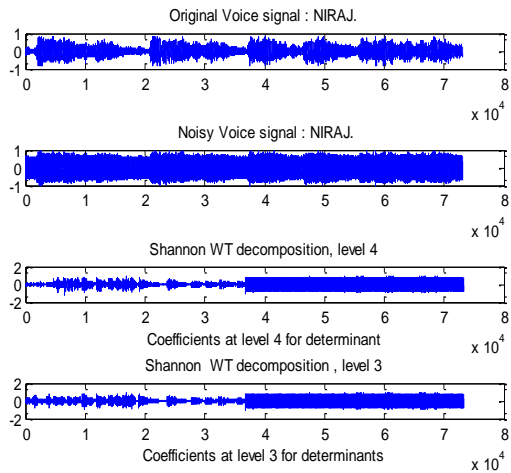

Fig. 6   Shannon WT Decomposition at stage- 4

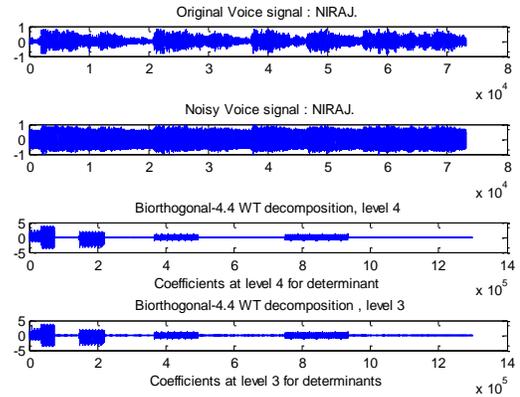

Fig. 7   Biorthogonal WT Decomposition at stage- 4

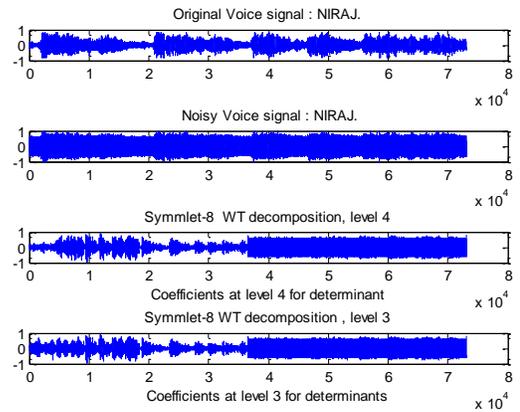

Fig. 8   Symmlet-8 WT Decomposition at stage- 4

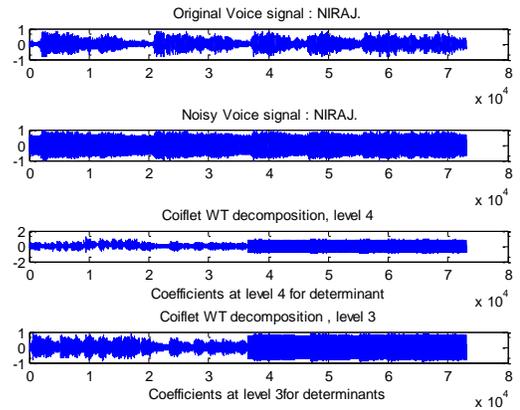

Fig. 9   Coiflet WT Decomposition at stage- 4

### 2.1.2 One Dimensional Signal Reconstruction (Synthesis) Stage-4 using Different Wavelets

1D signal reconstruction is done using a sequence of LP and HP filter banks at four different stages by cascading at LP upsample decimated by 2 as shown in Fig-10.

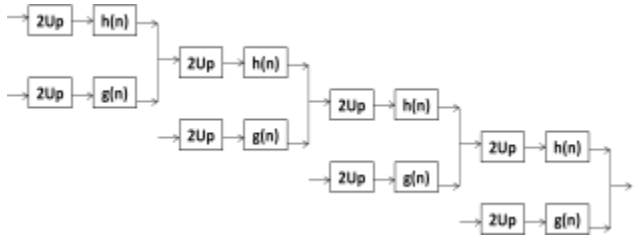

Fig. 10  1D signal Reconstruction at stage- 4

1D voice signal is reconstructed at stage-3 & stage-4 using DCT, FFT and different wavelets such as Haar, Deubachies, Morlet, Shannon, Biorthogonal, Symmlet and Coiflet as illustrated in Fig. 11 to Fig. 19.

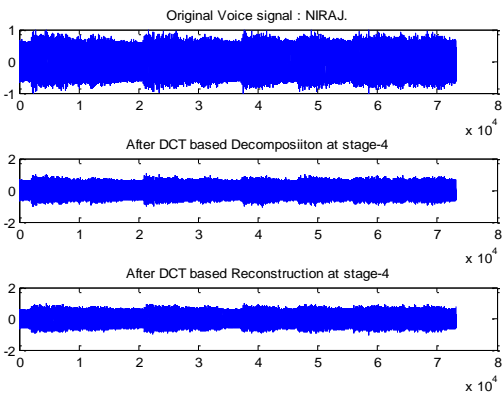

Fig. 11  DCT Analysis and Synthesis at stage-

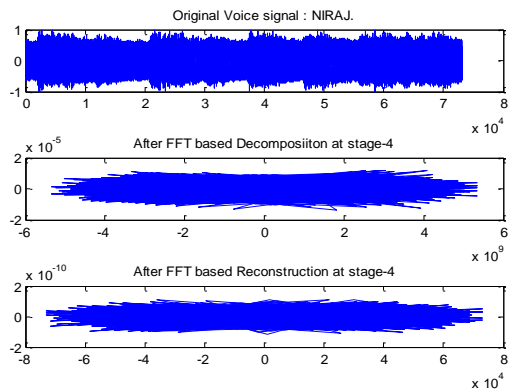

Fig. 12  FFT Analysis and Synthesis at stage- 4

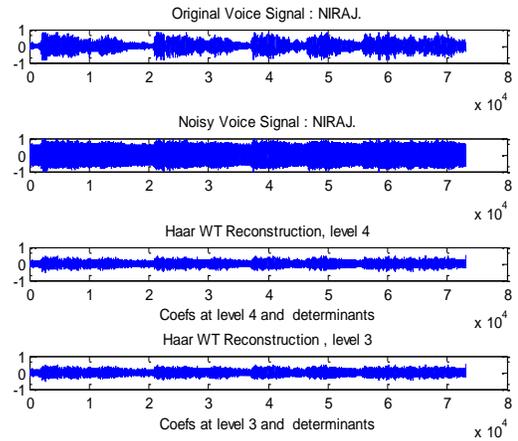

Fig. 13  Haar WT Reconstruction at stage- 4

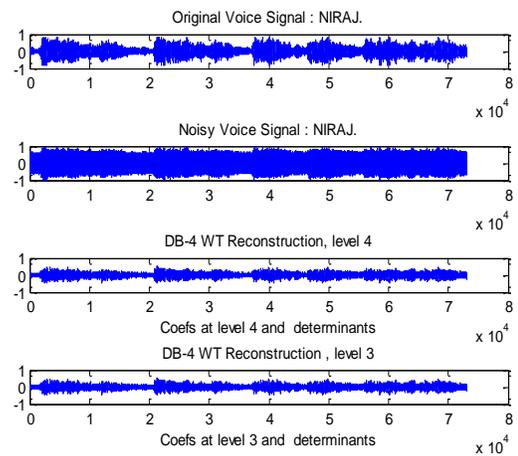

Fig. 14  DB-4 WT Reconstruction at stage- 4

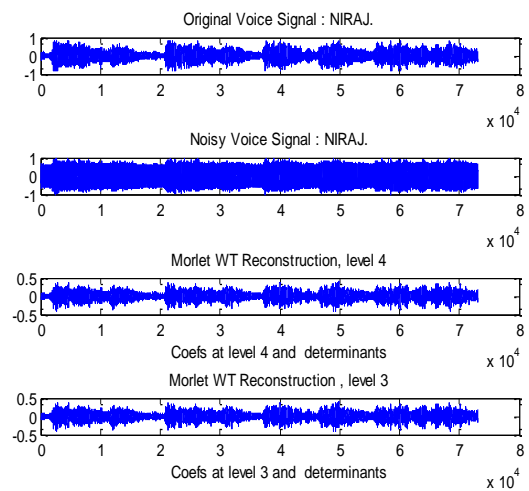

Fig. 15  Morlet WT Reconstruction at stage- 4

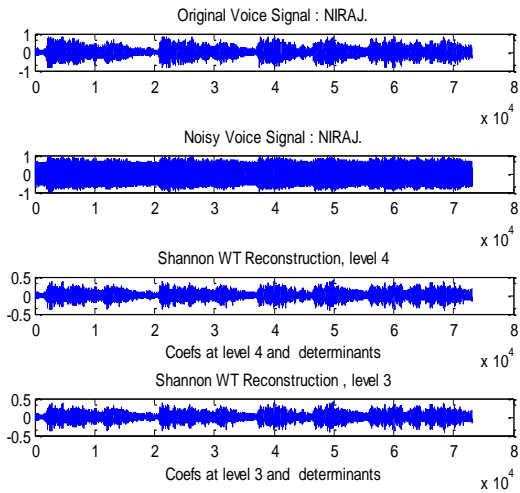

Fig. 16 Shannon WT Reconstruction at stage- 4

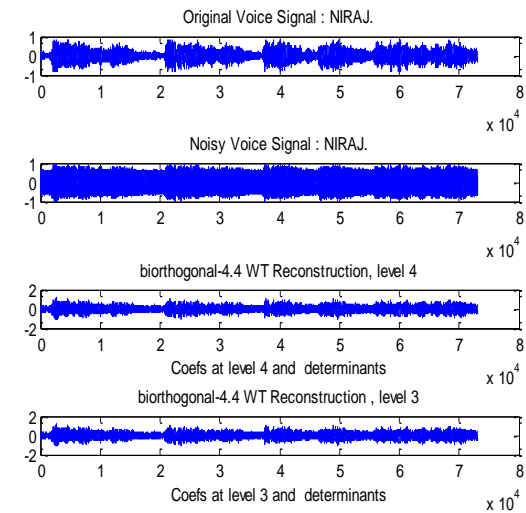

Fig. 17 Biorthogonal WT Reconstruction at stage-4

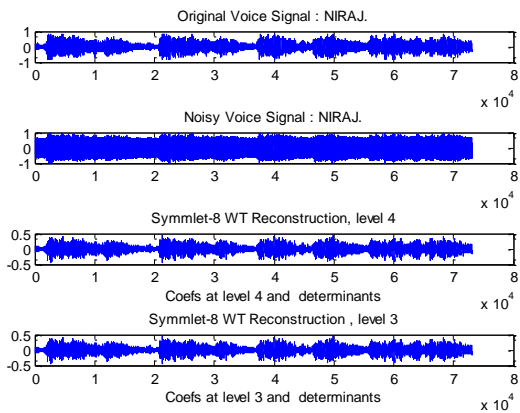

Fig. 18 Symmlet-8 WT Reconstruction at stage-4

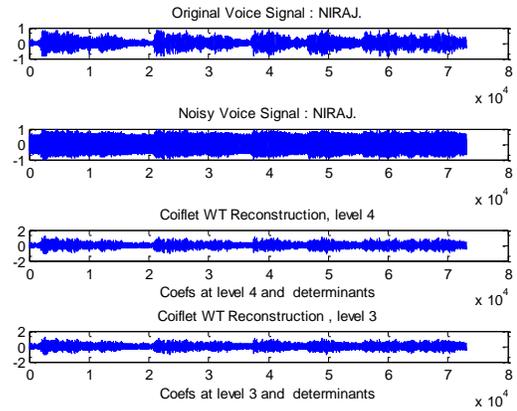

Fig. 19 Coiflet WT Reconstruction at stage- 4

### 2.1.3 Quantitative Analysis of Different Wavelets on 1D Noisy Voice Signal at stage-4

Table 1: Quantitative Analysis of Different Wavelets on Noisy Voice Signal at stage-4

| Different Wavelets | SNR (db) | MSE | Entropy | PSNR (db) |
|---|---|---|---|---|
| Haar | 81.39 | 1.4058e-034 | 4.1246 | 92.6011 |
| DB-4 | 60.06 | 1.9053e-012 | 3.8457 | 28.7194 |
| Morlet | 71.65 | 1.3231e-013 | 3.7496 | 17.1357 |
| Cauchy | 52.01 | 1.2170e-011 | 3.9384 | 36.7727 |
| Shannon | 74.20 | 7.3524e-014 | 3.7494 | 14.5841 |
| DCT | 24.69 | 6.5711e-009 | 4.7578 | 64.0962 |
| FFT | 97.27 | 1.0349e+004 | 4.6210 | 86.07 |
| Biorthogonal-2.4 | 72.67 | 1.9379e-006 | 4.4204 | 88.7932 |
| Biorthogonal-4.4 | 75.35 | 1.9358e-006 | 4.4407 | 88.7885 |
| Symmlet-8 | 67.54 | 3.4066e-013 | 3.7608 | 21.2431 |
| Coiflet | 52.01 | 1.2170e-011 | 3.9384 | 36.7727 |

From the quantitative analysis on voice signal at stage-4, it is found that Haar provides highest SNR, as well as lower MSE. Similarly, DCT and Bior-4.4 provide highest Entropy and Haar provides best PSNR. The average histogram is found approximately 7.3113e+003 in all cases. Specifically, Haar WT does not have overlapping windows, and reflects only changes between adjacent sample pairs. The Haar wavelet uses only two scaling and wavelet function coefficients, thus calculates pair wise averages and differences. That's why, Haar is found best

WT for noisy voice signal decomposition and reconstruction at stage-4.

### 2.2 Two Dimensional Image Decomposition and Reconstruction

2D discrete wavelet transform is used for 2D image decomposition and reconstruction using 2D scaling and wavelet functions which are orthogonal to its own translation [1-2], [5-6]. It consists of four sets of coefficients which are known as approximation coefficients, detail coefficients along the horizontal direction, detail coefficients along the vertical direction, detail coefficients along the diagonal direction.

#### 2.2.1 2D Image Analysis (Decomposition) at stage-4 using Different Wavelets

2D image decomposition is done using a sequence of combination of LP and HP filter banks in rows and columns (LL, LH, HL, HH) at four different stages by cascading at LL downsample decimated by 2 as shown in Fig. 20 and Fig. 21.

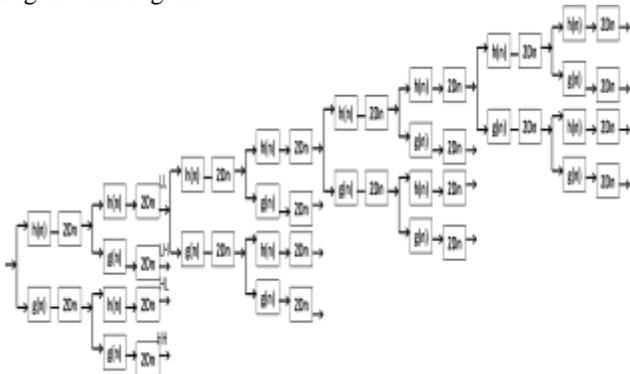

Fig. 20   2D Image Decomposition at stage- 4

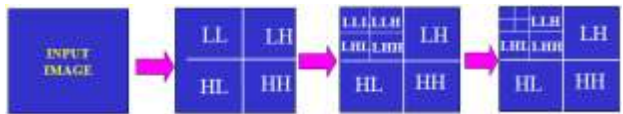

Fig. 21 Multiscale Analysis of 2D Image at stage-3

2D finger print image is decomposed at stage-2, stage-3 & stage-4 using DCT, FFT and different wavelets such as Haar, Deubachies, Morlet, Biorthogonal, Symmlet and Coiflet as illustrated in Fig.29, Fig.30 and Fig. 22 to Fig. 27.

#### 2.2.2 2D Image Synthesis (Reconstruction) at stage-4 using Different Wavelets Transform

2D image reconstruction is done using a sequence of combination of LP and HP filter banks in rows and columns (LL, LH, HL, HH) at four different stages by cascading at LL upsample decimated by 2 as shown in Fig. 28.

2D finger print image is reconstructed at stage-2, stage-3 & stage-4 using DCT, FFT and different wavelets such as Haar, Deubachies, Biorthogonal and Symmlet as illustrated in Fig. 29 to Fig. 35.

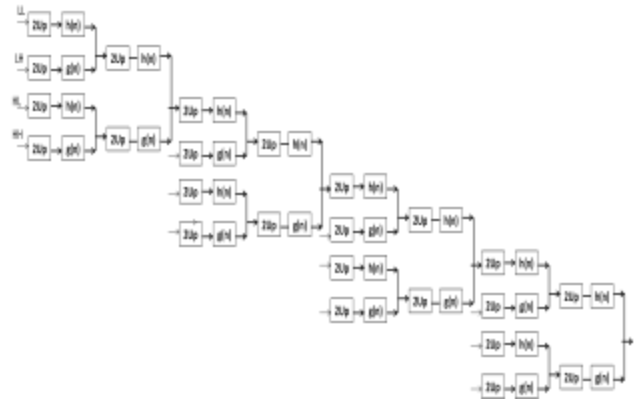

Fig. 28   2D Image Reconstruction at stage- 4

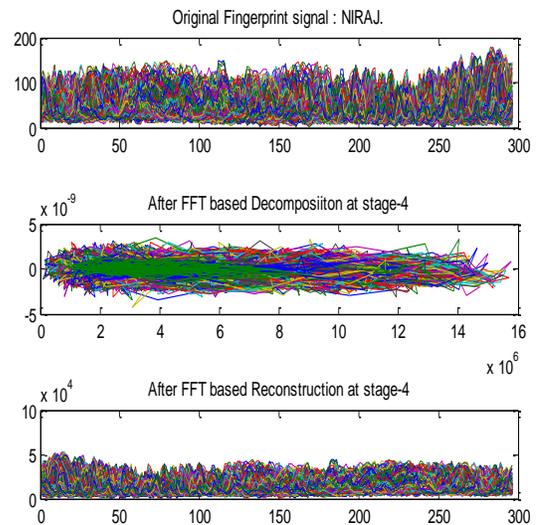

Fig. 29   FFT Analysis & Synthesis at stage -4

#### 2.2.3 Quantitative Analysis of Different Wavelets on 2D Image at stage-4

Table 2: Quantitative Analysis of Different Wavelets on 2D fingerprint image at stage-4

| Different Wavelets | SNR (db) | MSE | Entropy | PSNR (db) |
|---|---|---|---|---|
| Haar | 7.3167 | 8.4974 | 1.6483 | 28.03 |
| DB-4 | 8.4560 | 7.4169 | 1.9496 | 26.89 |
| Morlet | 8.9697 | 6.6969 | 1.7505 | 26.38 |
| DB-2 | 10.300 | 6.8423 | 2.0726 | 25.05 |
| Cauchy | 7.5217 | 6.3821 | 1.5782 | 21.03 |
| Shannon | 9.5147 | 6.6731 | 1.7338 | 25.83 |
| DCT | 7.9153 | 36.606 | 0.5591 | 26.13 |
| FFT-2 | 8.139 | 37.576 | 0.5129 | 25.36 |
| Biorthogonal-2.4 | 12.552 | 6.1138 | 1.9846 | 22.80 |
| Biorthogonal-4.4 | 12.448 | 6.3087 | 1.8981 | 22.90 |
| Symmlet-8 | 8.1908 | 7.3157 | 1.8272 | 27.16 |
| Coiflet | 10.300 | 6.8423 | 1.5726 | 25.05 |

From above quantitative analysis on 2D Fingerprint Image at stage-4, it is found that highest SNR and lowest MSE in Biorthogonal-2.4, higher MSE in Haar WT except FFT-2 & DCT. Similarly DB-2 as well as Bior-2.4 provides the highest Entropy and Haar as well as Sym-8 provides best PSNR. The average histogram is found 12.80 in all cases.

## 3. Two Dimensional Image Compression at stage-4

The salient compression performance is found 93.00% in DB-4, 93.68% in bior-4.4, 93.18% in Sym-4 and 92.20% in Coif-2 deploying hard threshold=30 at 4 stage analysis on the fingerprint image as illustrated in Tables 3-6.

Table 3: Compression performance using wavelets

| Daubechies | DB-4 | DB-6 | DB-8 | DB-10 |
|---|---|---|---|---|
| Compression | 93.00 | 92.21 | 91.04 | 90.08 |

Table 4: Compression performance using wavelets

| Biorthogonal | Bior2.4 | Bior4.4 |
|---|---|---|
| Compression | 91.47 | 93.68 |

Table 5: Compression performance using wavelets

| Symmlet | Sym4 | Sym8 | Sym10 | Sym25 |
|---|---|---|---|---|
| Compression | 93.18 | 91.53 | 90.49 | 84.49 |

Table 6: Compression performance using wavelets

| Different Wavelets | Coif2 | Coif5 | Rbio5.5 | Dmey |
|---|---|---|---|---|
| Compression | 92.20 | 88.34 | 89.70 | 76.59 |

Daubechies wavelets are seen compactly supported and have highest number of vanishing moments whereas Biorthogonal wavelets compactly supported wavelets for symmetry and exact reconstruction. The time taken is not linear for wavelet decomposition and reconstruction. On the other hand, Symmlets are compactly supported wavelets with highest number of vanishing moments and the best compression is given by Sym-8 [6-8].

The original image and compressed image deploying different wavelets such as Sym-8, Bior-2.4 and Bior-4.4 and Db-4 are illustrated in Fig. 36 and Fig. 37.

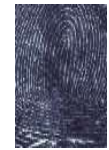

Uncompressed Fingerprint Image

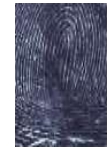

20-to-1 Compression, bior-4.4 Wavelet, Threshold = 30, level=4

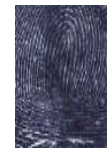

20-to-1 Compression, bior-2.4 Wavelet, Threshold = 30, level=4

Fig. 36 Image compression by Bior- 4.4 & Bior- 2.4

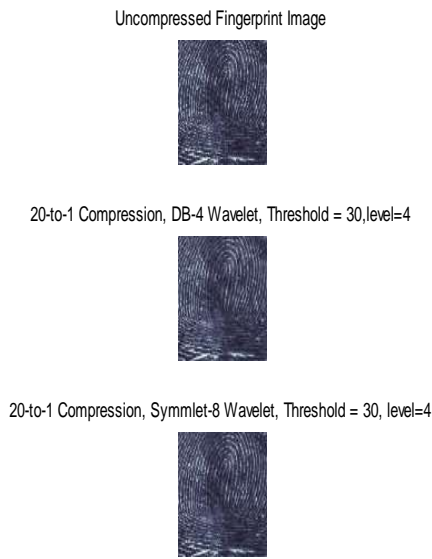

Fig. 37 Image compression using DB-4 and Sym-8

## 4. Conclusion

Multiscale analysis concludes that MSE is increased with the increasing number of stages whereas the SNR decreases. The quantitative analysis on 1D noisy voice signal at stage-4, shows that highest SNR and lower MSE in Haar, highest Entropy in DCT and Bior-4.4 and best PSNR in Haar. On the other hand, the quantitative analysis in 2D fingerprint image at stage-4, concurs that highest SNR and lowest MSE in Biorthogonal-2.4, higher MSE in Haar WT except FFT-2 & DCT. Similarly DB-2 as well as Bior-2.4 provides highest Entropy and Haar as well as Sym-8 provides best PSNR. Furthermore, the simulation results show the significant image compression. The salient compression performance is found that 93.00% in DB-4, 93.68% in bior-4.4, 93.18% in Sym-4 and 92.20% in Coif-2 deploying hard threshold=30 at stage 4 on the 2D fingerprint image. Future research will concentrate the application of next generation of wavelets on video frames.

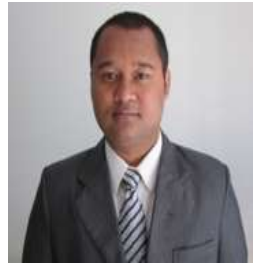

**Dr. Niraj Shakhakarmi** worked as a Doctoral researcher from 2009 to 2011 in the US ARO (Army Research Office) funded Center for Battlefield Communications (CeBCom) Research, Department of Electrical and Computer Engineering, Prairie View A&M University (Texas A&M University System). He received his B.E. degree in Computer Engineering in 2005 and M.Sc. in Information and Communications Engineering in 2007. He has accomplished Ph.D in Electrical & Computer Engineering in 2011 from Prairie View A&M University, Houston, USA. His research interests are in the areas of Wavelets applications and Digital Image Processing, Secured Position Location & Tracking (PL&T), Cognitive Radio Networks, Mobile Ad hoc Networks, 4G Networks, Satellite Networks, QoS & Network Security, and Color Technology. He has published WSEAS journal, ICSST conference, Elsevier conference, and several papers are under review in IEEE and WSEAS journals and conference papers.